# A Novel Quantum Cost Efficient Reversible Full Adder Gate in Nanotechnology

Md. Saiful Islam

*Institute of Information Technology, University of Dhaka, Dhaka-1000, Bangladesh*

E-mail: saiful@iit.du.ac.bd

## Abstract

*Reversible logic has become one of the promising research directions in low power dissipating circuit design in the past few years and has found its applications in low power CMOS design, cryptography, optical information processing and nanotechnology. This paper presents a novel and quantum cost efficient reversible full adder gate in nanotechnology. This gate can work singly as a reversible full adder unit and requires only one clock cycle. The proposed gate is a universal gate in the sense that it can be used to synthesize any arbitrary Boolean functions. It has been demonstrated that the hardware complexity offered by the proposed gate is less than the existing counterparts. The proposed reversible full adder gate also adheres to the theoretical minimum established by the researchers.*

## 1. Introduction

Power dissipation is one of the most important factors in VLSI circuit design. Irreversible logic circuits dissipates kT*log 2 Joule (k is the Boltzmann constant and T is the absolute temperature) heat for every bit of information that is lost irrespective of their implementation technologies [1]. Information is lost when the input vectors cannot be recovered from circuit's output vectors. Reversible logic naturally takes care of heating since in reversible circuits the input vectors can be uniquely recovered from its corresponding output vectors. Bennett [2] showed that zero energy dissipation is possible only if the gating network consists of reversible gates. Thus reversibility will become future trends towards low power dissipating circuit design.

  Reversible logic design differs significantly from traditional combinational logic design approaches. In reversible logic circuit the number of input lines must be equal the number of output lines, each output will be used only once and the resulting circuit must be acyclic [3]. The output lines that are not used further are termed as garbage outputs. One of the most challenging tasks is to reduce these garbages [3]. Any reversible logic gate realizes only the functions that are reversible. But many of the Boolean functions are not reversible. Before realizing these functions, we need to transform those irreversible functions into reversible one. Any transformation algorithm that converts an irreversible function to a reversible one introduces input lines that are set to zero in the circuit's input side [4-5]. These inputs are termed as constant inputs. Therefore, any efficient reversible logic design should minimize the garbages as well as constant inputs.

  This paper presents a novel 4*4 reversible gate namely Peres Full Adder Gate (PFAG), that is, it has 4-input lines and 4-output lines. This gate can be used to realize any arbitrary Boolean function and therefore universal. The hardware complexity of this gate is less compared to the existing ones and requires only one clock cycle. The quantum realization cost of this gate is only 8 and ready for use in current nanotechnology.

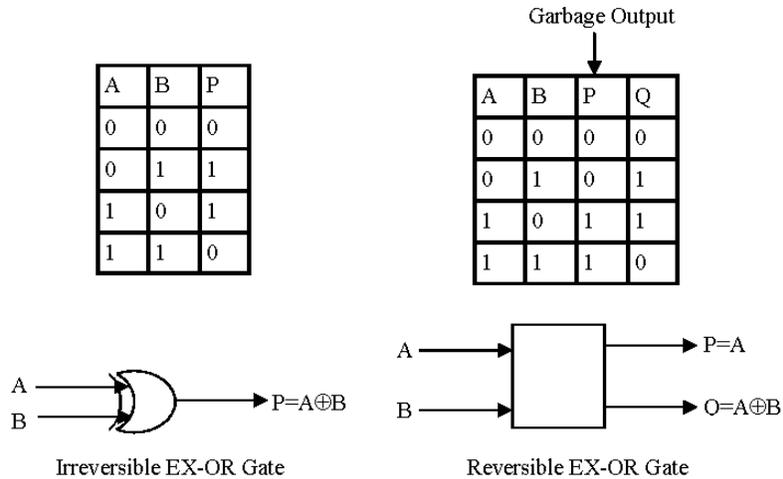

Figure 1: Irreversible and reversible gate

## 2. Reversible Logic Gates

There exist many reversible gates in the literature. Among them 2*2 Feynman gate [6] (shown in figure2), 3*3 Fredkin gate [7] (shown in figure 3), 3*3 Toffoli gate [8] (shown in figure 4) and 3*3 Peres gate [9] (shown in figure 5) is the most referred. The detailed cost of a reversible gate depends on any particular realization of quantum logic. Generally, the cost is calculated as a total sum of 2*2 quantum primitives used. The cost of Toffoli gate is exactly the same as the cost of Fredkin gate and is 5. The only cheapest quantum realization of a complete (universal) 3*3 reversible gate is Peres gate and its cost is 4.

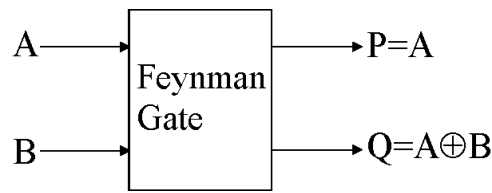

Figure 2: 2*2 Feynman gate

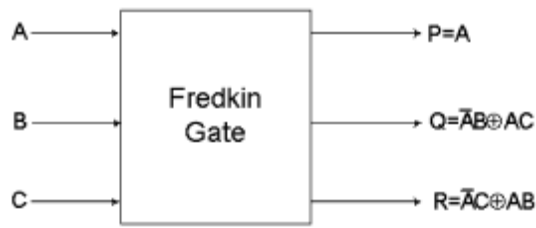

Figure 3: 3*3 Fredkin gate

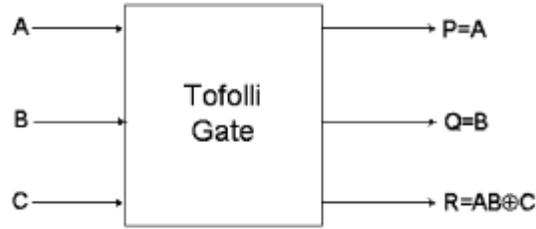

Figure 4: 3*3 Toffoli gate

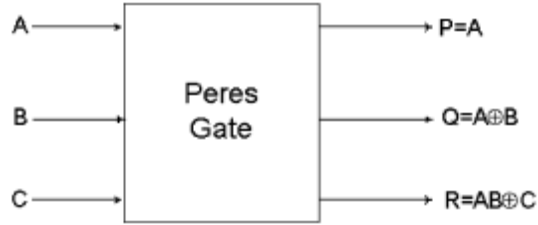

Figure 5: 3*3 Peres gate

## 3. Reversible Logic Implementation of Full Adder Circuit

Full adder is the fundamental building block in many computational units. The anticipated paradigm shift logic compatible with optical and quantum requires compatible reversible adder implementations. The full adder circuit's output is given by the following equations:

$$Sum = A \oplus B \oplus C_{in}$$

$$C_{out} = (A \oplus B)C_{in} \oplus AB$$

The reversible logic implementation of full-adder circuit and other adder circuits and their minimization issues has been discussed in [10-13]. It has been shown in [11] and [13] that any reversible logic realization of full adder circuit includes at least two garbage outputs and one constant input. The author in [10-13] has given a quantum cost efficient reversible full adder circuit that is realized using two 3*3 Peres gates only (shown in figure 6). This implementation of reversible full adder circuit is also efficient in terms of gate count, garbage outputs and constant input than the existing counter parts.

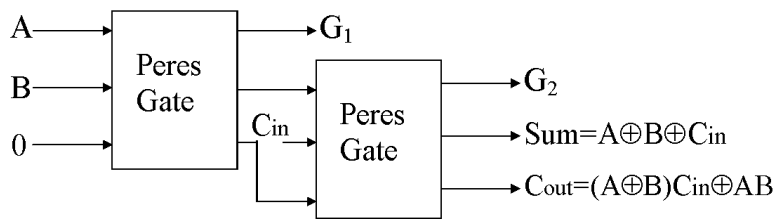

Figure 6: The only cheapest quantum realization of reversible full adder circuit [10-13]

## 4. A Novel Reversible Full Adder Gate

Full adder is the fundamental building block in almost every arithmetic logic circuit. Therefore, a gate that can work singly as a reversible full adder will be beneficial to the development of other complex logic circuits. This paper presents a novel reversible full adder gate namely Peres Full Adder Gate (PFAG) shown in figure 7. The gate is achieved by cascading two 3*3 Peres gate. The quantum realization cost of this gate is 8 (shown in figure 8) since it includes two 3*3 Peres gates. The gate can work singly as a reversible full adder circuit when its fourth input is set to zero (D=0) as shown in figure 9. This gate requires only one clock cycle and produces no extra garbage outputs, that is, it adheres to the theoretical minimum as established in [11] and [13].

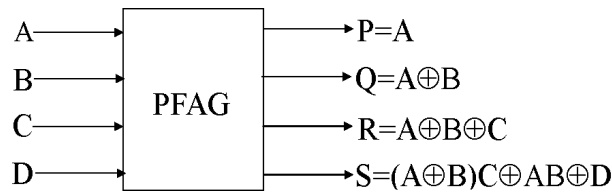

Figure 7: 4*4 Peres Full Adder Gate

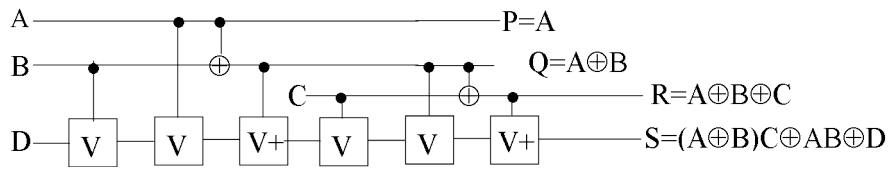

Figure 8: Quantum Realization of Peres Full Adder Gate

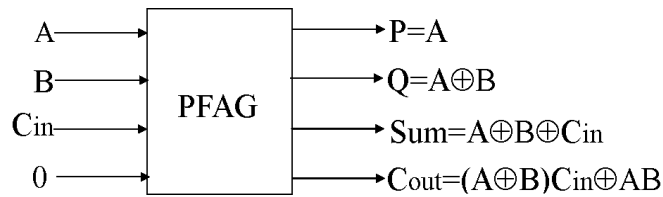

Figure 9: PFAG gate as a reversible full adder circuit

The proposed reversible full adder gate is not the only one that can work singly as a reversible full adder circuit. There are other three reversible full adder gates in the literature namely TSG [14] (shown in figure 10), HNG [15] (shown in figure 11), and MKG [16] (shown in figure 12) exists in the literature. The functionality of HNG and PFAG is almost similar except that PFAG provides half adder sum output besides of giving full adder sum output. This extra output will be advantageous to the development of other adder circuit such as carry skip adder [13] and BCD adder. The quantum realization of TSG, HNG and MKG is not given in the literatures so far and is therefore unknown.

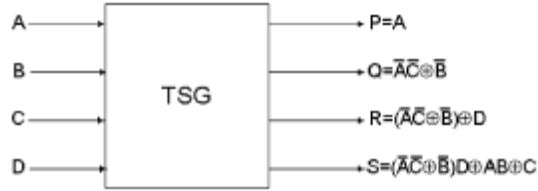

Figure 10: 4*4 TSG gate

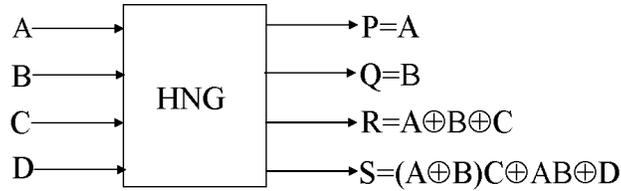

Figure 11: 4*4 HNG gate

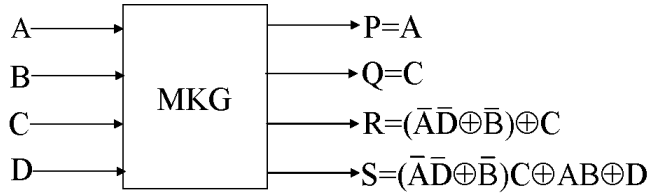

Figure 12: 4*4 MKG gate

5. **Results and Discussion**

The following demonstrates that the proposed reversible full adder gate is superior to the existing counterparts in terms of hardware complexity, quantum costs, garbage outputs and constant inputs.

**Hardware Complexity:** One of the main factors of a gate is its hardware complexity. It can be proved that proposed reversible PFAG gate better than the existing counterparts in terms of hardware complexity. Let

$\alpha$ = A two input EX-OR gate calculation

$\beta$ = A two input AND gate calculation

$\delta$ = A NOT gate calculation

T = Total logical calculation

For TSG the total logical calculation T is $6\alpha+3\beta+3\delta$. For HNG total logical calculation T is $5\alpha+2\beta$. For MKG total logical calculation T is $5\alpha+3\beta+3\delta$. For PFAG total logical calculation T is $5\alpha+2\beta$ (when D is set to zero it requires five two input EX-OR gate calculations). Thus the proposed reversible full adder gate is better than the existing ones in terms of hardware complexity. In addition of it its second output can be used as the half adder sum output.

**Quantum Costs:** The detailed cost of a reversible gate depends on any particular realization of quantum logic. The quantum realization cost of the proposed reversible full adder gate is only 8. This realization cost is equal to the cost of the cheapest quantum realization of reversible full adder circuit given in [10-13]. The quantum realization cost of TSG, HNG, and MKG is not given by the authors and is therefore unknown now.

**Garbage Outputs:** The proposed reversible full adder gate does not produce any extra garbage outputs. The proposed gate produces only 2 garbage outputs and it is theoretically minimum as mentioned in [10-13].

**Constant Inputs:** The proposed reversible full adder gate requires only one constant input and it is theoretically minimum as mentioned in [10-13].

Table 1: Comparative experimental results of different reversible full adder gate

| Reversible Full Adder Gate | Two Input EX-OR Gate Calculation ($\alpha$) | Two Input AND Gate Calculation ($\beta$) | NOT Gate calculation ($\delta$) | Total Logical Calculation (T) | Quantum Costs |
|---|---|---|---|---|---|
| **Proposed PFAG** | 5 | 2 | 0 | $5\alpha+2\beta$ | 8 |
| **HNG[15]** | 5 | 2 | 0 | $5\alpha+2\beta$ | Unknown |
| **MKG [16]** | 5 | 3 | 3 | $5\alpha+3\beta+3\delta$ | Unknown |
| **TSG [14]** | 6 | 3 | 3 | $6\alpha+3\beta+3\delta$ | Unknown |

6. **Conclusion**

This paper presents a new quantum cost efficient reversible full adder gate in nanotechnology. This gate requires only clock cycle and can be used to synthesize any arbitrary Boolean functions therefore universal. The hardware complexity offered by this gate is less than the existing reversible full adder gates. The quantum realization cost of this gate is only 8. This gate is readily available for use in nanotechnology since its quantum implementation is given in NMR technology [13].